\documentclass[aps, nofootinbib]{revtex4}
\usepackage{amsmath, amssymb, bm, graphicx, graphics, color,mathrsfs,hyperref}

\newcommand  {\Rbar} {{\mbox{\rm$\mbox{I}\!\mbox{R}$}}}

\newcommand{\Lie}[0]{{\cal L}\, }

\newcommand{\tl}{\theta_{(\ell)}}
\newcommand{\tn}{\theta_{(n)}}

\newcommand{\nn}{\nonumber}
\newcommand{\be}{\begin{equation}}
\newcommand{\ee}{\end{equation}}
\newcommand{\bea}{\begin{eqnarray}}
\newcommand{\eea}{\end{eqnarray}}

\newcommand{\tq}{\tilde{q}}
\newcommand{\hu}{\hat{u}}

\newcommand{\tom}{\tilde{\omega}}

\newcommand{\cV}{\mathcal{V}}

\begin{document}

\title{Spacetimes containing slowly evolving horizons}
\date{\today}
\author{William Kavanagh}
\email{wkavanag@physics.mun.ca}
\affiliation{Department of Physics and Physical Oceanography, Memorial University of Newfoundland\\
St. John's, Newfoundland and Labrador, A1B 3X7, Canada}
\author {Ivan Booth}
\email{ibooth@math.mun.ca}
\affiliation{Department of Mathematics and Statistics, Memorial
 University of Newfoundland \\  
St. John's, Newfoundland and Labrador, A1C 5S7, Canada}

\begin{abstract}
Slowly evolving horizons are trapping horizons that are ``almost" isolated horizons. This paper 
reviews their definition and discusses several spacetimes containing such structures. These include 
certain Vaidya and Tolman-Bondi solutions as well as (perturbatively) tidally distorted black holes. 
Taking into account the mass scales and orders of magnitude that arise in these calculations, 
we conjecture that slowly evolving horizons are the norm rather than the exception in 
astrophysical processes that involve stellar-scale black holes. 
\end{abstract}

\maketitle

\section{Introduction}

Physics is most exciting far from equilibrium. Few would argue that laminar fluid flow is more
interesting than turbulence or that a slowly cooling cup of tea is more intriguing
than the 
rapidly boiling kettle of water that directly preceded it. That said, it is precisely the relative dullness of 
equilibrium or near-equilibrium systems that helps us to understand them so well. 
In turn, this understanding is the foundation on which much of our knowledge of nature is based. 

Though black holes are intrinsically extreme objects, they too have equilibrium states. Traditionally 
these have been identified with the Kerr family of spacetimes which consists of non-evolving but rotating
black holes sitting alone in an otherwise empty, asymptotically flat universe. Given that these are perturbatively stable and have been shown to be the unique stationary, rotationally symmetric and asymptotically flat vacuum black hole spacetimes,  it is widely believed that in the absence of external 
interactions all black holes will ultimately settle down into a state that is closely approximated by one
of these solutions (see, for example, \cite{wald} for a discussion of these results). 
Much of what we know about black holes 
comes from a study of these equilibrium solutions and their perturbations. 

Of course, a key word in the above paragraph is ``approximated". Real astrophysical black holes 
don't each occupy their own asymptotically flat universe.  Thus when considering a real black hole, one
typically assumes that the rest of the universe has only a weak influence on the spacetime 
structure close to the black hole.  Then, in that neighbourhood the spacetime should
 be perturbatively Kerr. 
%
%
In the last decade an alternative approach to this problem has been proposed. The \emph{isolated horizon} programme (see \cite{isoreview, LivRev,gourgoul} for reviews) explicitly studies the quasilocal conditions under which a  black hole may be said to be in equilibrium with its (possibly dynamic) surroundings and then studies consequences of those conditions -- as opposed to studying equilibrium spacetimes which also contain black holes.  

Such a characterization is important if one wishes to consider 
cases where a black hole may sometimes dynamically interact with its surroundings and at 
other times be quiescent. While the standard causal definitions of black holes and event horizons work 
well for stationary spacetimes, they encounter serious problems for these more general situations. A
primary problem is that dynamic, causally defined black holes cannot even be precisely located 
by non-omniscient observers. However, there are secondary problems as well. A particularly serious 
one for those interested in equilibrium states follows from the area increase theorem. This says that an 
event horizon can never decrease in area and what is more its rate of expansion must always 
decrease. Thus, due to their teleological definition, event horizons will grow in anticipation of 
future interactions even when there is no proximate cause for such an expansion. In particular,
lack of growth cannot be used to characterize (temporary) equilibrium states (see \cite{BHboundaries}
for a further discussion of these points). 


Following these arguments for quasilocal definitions,
Booth and Fairhurst \cite{prl}  recently 
proposed a characterization of near-equilibrium
black holes : \emph{slowly evolving horizons}. These ``almost isolated" horizons include 
(most) truly isolated horizons as particular examples and in turn are special cases of Hayward's 
\emph{trapping horizons} \cite{hayward} and are closely related to standard apparent horizons 
as well as 
Ashtekar and Krishnan's \emph{dynamical horizons} \cite{ak, LivRev, SKB} (see \cite{BHboundaries} for a detailed discussion of how these objects are related). 

In this paper, we shall consider several examples of spacetimes containing 
slowly evolving horizons including certain
Vaidya (section \ref{vaidya}) and Tolman-Bondi (section \ref{TB}) spacetimes as well as 
a tidally perturbed Schwarzschild solution (section \ref{tidal}). In these examples we will see that 
the spacetimes that we would intuitively expect to be near equilibrium do indeed contain slowly
evolving horizons. However we will also see that some fairly extreme situations also turn out to be 
slowly evolving. For example, we will find that even the formation of a black hole during gravitational 
collapse can be a slowly evolving process.

\section{Preliminaries}

We begin with a review of the definition of a slowly evolving horizon (section \ref{SEHdef}) and a brief
consideration of the units that we will use in our examples (section \ref{units}).

\subsection{Slowly evolving horizons}
\label{SEHdef}

A few preliminary definitions are needed before we consider that of a slowly evolving horizon proper. 
First, a \emph{marginally trapped surface} (MTS) is a closed two-surface with $\tl = 0$ and $\tn < 0$ \cite{gregabhay}. Here $\tl = \tq^{ab} \nabla_a \ell_b$ and 
$\tn = \tq^{ab} \nabla_a n_b$ are respectively the expansions of the outward ($\ell$) and inward
($n$) forward-in-time pointing null normal vectors to the two-surface and $\tq_{ab} = g_{ab} + \ell_a n_b + n_a \ell_b$ 
is the transverse two-metric on that surface. Throughout this paper we will cross-normalize the null vectors
so that $\ell \cdot n = -1$ and tacitly assume that all MTSs are topologically $S^2$. 

Then, a \emph{marginally trapped tube} (MTT)  is a three-surface of arbitrary signature that can be entirely foliated by MTSs \cite{ak} and a \emph{future outer trapping horizon} (FOTH) is an MTT which also
satisfies $\delta_n \tl < 0$ \cite{hayward}. Here $\delta_n$ refers to a deformation  
generated by $n$ \cite{BHboundaries}  
and so the condition says that each MTS can be deformed into a surface with $\tl < 0$ 
by an arbitrarily small evolution ``inwards". For a sufficiently small deformation $\tn$ must also remain negative, and so this condition says that there are fully trapped surfaces ``just inside" the MTT and untrapped ones outside. Thus a FOTH
can be regarded as a three-dimensional generalization of an apparent horizon \footnote{\label{one} 
This intuitive
picture must be qualified. It is well-known that apparent horizons are slicing dependent. Similarly
FOTHs are not uniquely defined and may themselves be smoothly deformed. Thus, while
it is certainly true that a given MTT separates an associated set of trapped from untrapped surfaces,
it will also be true that other equally valid MTTs and even full trapped surfaces will intersect it
(though in restricted ways). These issues are still under active investigation. See, for example, 
\cite{eardley, hayward, gregabhay, ams, BHboundaries,SchnetKrish}.}.


In defining a slowly evolving horizon, it is convenient to further restrict the scaling of the null vectors. 
To do this we label the foliating MTSs with a parameter $v$ and choose the 
scaling and an \emph{evolution parameter} $C$  so that
\be
\mathcal{\cV}^a = \ell^a - C n^a \label{cV} 
\ee
is tangent to the MTT and 
\be 
\Lie_{\cV} v = 1 \label{C} \, . 
\ee
Thus $\cV$ evolves the MTSs into each other and in particular the $C$ characterizes how the
area element $\sqrt{\tq}$ on the two-surfaces changes with increasing $v$. We have
\be 
\Lie_{\cV} \sqrt{\tq}= - C \tn \sqrt{\tq} \label{Adot}
\ee
and so the area increases if $C>0$ (the horizon is spacelike), stays the same if $C=0$ (the horizon is null), or decreases if $C<0$ (the horizon is timelike). 

The value of the expansion parameter 
is directly determined by the behaviour of the matter and gravitational fields at the 
horizon surface. It was first shown in \cite{hayward} that if the null energy condition
holds, a FOTH expands if the shear $\sigma^{(\ell)}_{ab} = \tq_a^c \tq_b^d \nabla_c \ell_d$ and/or
the matter flux $T_{ab} \ell^a \ell^b$ are non-zero. That is, it is spacelike and expanding 
if there is a flux of matter or gravitational energy across the horizon. 
Otherwise, it is null and so is an isolated horizon. This may be thought of as the second law for
FOTHs and in fact it has recently been sharpened : if either
$\sigma^{(\ell)}_{ab} \neq 0$ or $T_{ab} \ell^a \ell^b \neq 0$ anywhere on a particular MTS of a FOTH, then one can apply a maximal principle to the partial differential equation for $C$ defined by $\delta_{\cV} \tl = 0$ to show that $C>0$ everywhere on that two-surface. By the same reasoning, 
if $C=0$ anywhere on an MTS, then it is zero everywhere. 
Thus, there are no ``partially isolated" MTSs and each
MTS is either expanding everywhere or expanding nowhere \cite{ams, BHboundaries, defPaper}. 
One consequence of this is that a FOTH can be cleanly split into isolated and non-isolated 
regions and the boundaries between these regions are two-surfaces of the foliation. 


We finally come to our definition \cite{prl,defPaper} : 
\\ \\
\textbf{Definition:} A region of a future outer trapping horizon $H$ with $v \in [v_1, v_2]$ is a 
\emph{slowly evolving horizon} (SEH) \emph{of order $\epsilon$} if
\newcounter{ctr}
\begin{list}{S\arabic{ctr}.}{}
\usecounter{ctr}
\item $\epsilon = \sqrt{C \tn^2} R_H \ll1$ ,  \label{s1}
\item the null vectors are scaled so that $| \cV | = \sqrt{2 C} \sim \epsilon$, \label{s2}
\item  $| \Lie_{\cV} \tom_a |  \lesssim \epsilon/R_H^2$ and $|\Lie_{\cV} \tn | \lesssim \epsilon/R_H^2$ and \label{s3}
\item $\tilde{\mathcal{R}}$, $|\tom|^2$, $|\sigma^{(n)}|^2$, and $T_{ab} n^a n^b \sim 1/ R_H^2$ or smaller. 
\label{s4}
\end{list}
In the above $R_H(v)$ is the areal radius of the corresponding MTS, $\tilde{\mathcal{R}}$ is the two-dimensional 
Ricci scalar on those surfaces, $\tom_a = - \tq_a^b n_c \nabla_b \ell^c$ is the well-known angular mometum one-form (or equivalently the connection on the normal bundle), and $\sigma^{(n)}_{ab}$ is the shear associated
with the ingoing null direction $n$. 

Very briefly, these conditions may be interpreted in the following way. First note that even though 
both $C$ and $\tn$ explicitly depend on the scaling of the null vectors, 
the combination $C \tn^2$ does not. For an isolated horizon it vanishes, while 
if $H$ is spacelike we have
$
\Lie_{\hat{\cV}} \sqrt{\tq} = - \sqrt{{C}/{2}} \tn \sqrt{\tq} 
$, 
where $\hat{\cV} = \cV/\sqrt{2C}$. 
Thus $\epsilon^2 = C \tn^2 R^2_H$ is roughly the square of a normalized rate of expansion of the area element and clearly independent of the foliation parameter $v$. 
Condition S\ref{s1} is the key defining condition for slowly evolving horizons and in the 
upcoming examples we will often refer to $\epsilon$ as the \emph{slowly evolving parameter}.

The second condition (S\ref{s2}) is not a geometric invariant of the surface. 
Instead it is a matter of picking a convenient labelling for the surfaces (equivalently 
a scaling of the null vectors) to reflect their slowly evolving nature. One way to ensure
that this condition holds is to scale the null vectors so that $\tn \sim 1/R_H$. In particular for
all of our examples we will choose $\tn = 2/R_H$ (which is the value that it takes for Schwarzschild 
spacetimes with a standard scaling of $n$).  

One of the main reasons for 
choosing the foliation parameter in this way is to make the last two conditions easier to 
state. Then, condition S\ref{s3} says that the angular momentum one-form and $\tn$ also change slowly with respect to such a parameter, while condition S\ref{s4} requires that conditions in a neighbourhood of the horizon not be too extreme.

On applying these conditions to the evolution and embedding equations for FOTHs,
it can be shown that an SEH also obeys dynamic zeroth and first laws of black hole mechanics. 
If one defines the surface gravity of the horizon by 
\be
\kappa_{\cV} = - \cV^a n_b \nabla_a \ell^b
\ee
in analogy with the definition 
for event, Killing, or isolated horizons then
\be
\kappa_{\cV} =  \kappa^{(0)} + \epsilon \kappa^{(1)} + \mbox{O}(\epsilon^2)  \label{zeroth} 
\ee
for some constant $\kappa^{(0)}$ and function $\kappa^{(1)}$. Over any isolated segment of the 
horizon $\epsilon = 0$ and one recovers the usual zeroth law. However, even when the horizon isn't strictly isolated, we see that the surface gravity only varies by order $\epsilon$ 
across any MTS and, in fact, 
between MTSs during evolutions. This is what one would expect for a quasi-equilibrium state. 

The restrictions imposed by the slowly evolving conditions are not strong enough to fix a unique 
value for $\kappa^{(0)}$. However, at least for spherical symmetry, it is fairly easy to see that a natural
scaling choice recovers the usual Schwarzschild value. This follows from the embedding
equations which tell us that along a spherically symmetric MTT \cite{defPaper} :  
\bea
 \Lie_{\cV} \tn  
 & = & 
  -  \kappa_{\cV} \tn - \tilde{\mathcal{R}}/2 + 8 \pi T_{ab} \ell^a n^b 
  + C \left(8 \pi T_{ab} n^a n^b + \tn^2/2 \right) \, .  
\eea
Then if we scale the null vectors so that $\tn = -2 / R_H$, we have 
\bea
\kappa_{\cV} = \frac{1}{2 R_H} \left(1 + 8 \pi R_H^2 T_{ab} (\ell^a + C n^b) n^b \right)  \, . \label{kap0}
\eea
Thus, if $T_{ab} = 0$ (as is the case for Schwarzschild spacetime) 
we find that $\kappa = 1/2R_H$. However
even if it doesn't vanish completely, it is clear that we will have 
$\kappa^{(0)} = 1/2R_H$ if $T_{ab} (\ell^a + C n^b) n^b$ is small relative to the size of 
the MTS. In our examples we will see that this is the case when the horizons are slowly 
evolving. 

That said, in many cases such a ``correct" scaling will not be obvious and then it is important to keep in mind that the SEH will still obey a zeroth law no matter what the choice. 
Further it can be shown that to second order in $\epsilon$ it will also obey a first law : 
\bea
\frac{1}{8 \pi G} \kappa^{(0)} \frac{da_H}{dv} \approx \int_{H_v} d^2 x \sqrt{\tq} \left\{ T_{ab} \ell^a \ell^b 
+\frac{1}{8 \pi G} |\sigma^{(\ell)}|^2 \right\}   \, , \label{firstLaw}
\eea
where $H_v$ is a foliating MTS and $a_H(v)$ is the area of that surface. This result recalls 
the standard physical process
versions of the first law \cite{gao}, the first law for dynamic event horizons \cite{hawking} and the
flux laws for dynamical horizons \cite{ak}. 
 
In cases where these horizons have an (approximate) axis of symmetry, it also makes sense to define their angular momentum and this can be incorporated into the first law in much the way one would expect. However as our examples will all be either exactly or 
perturbatively spherically symmetric, we will not consider the angular momentum terms in much 
detail here. 

A more detailed discussion of all of these matters may be found in \cite{prl, defPaper}.

\subsection{Units}
\label{units}
To build an intuition as to when a black hole will be slowly evolving we will often need to translate our 
examples into physical units. Since we are primarily interested in astrophysical processes, we take our standard unit of measurement to be one solar mass :
\be
M_\odot = 1.9 \times 10^{30} \, \mbox{kg} \, .
\ee
Then, using the gravitational constant 
$G=6.7 \times 10^{-11} \, \mbox{m}^3 \, \mbox{kg}^{-1} \, \mbox{s}^{-2}$
and the speed of light $c = 3.0 \times 10^8\,  \mbox{m} \,  \mbox{s}^{-1}$
distances and times that come out of our equations will be measured in 
units
\bea
R_{\odot} &=& GM_\odot/c^2 \approx 1.4 \times 10^3 \,  \mbox{m}  \, \, \, \mbox{and} \\
T_\odot & = & R_\odot/c \approx 4.7 \times 10^{-6} \, \mbox{s} \,  
\eea
respectively. 

In some of the examples we will consider dust and have expressions for its
density. Then, it will be useful to have a few density reference points for comparison. Note that 
\bea
\rho_\odot  & \approx & 1.3 \times 10^3 \mbox{kg}/\mbox{m}^3 
\approx 1.9 \times 10^{-18} /M_\odot^2 \, , \\
\rho_{\mbox{wd}}  
& \approx & 10^9 \mbox{kg}/\mbox{m}^3 \approx 1.4 \times 10^{-12} /M_\odot^2 \,  \, \, \, \mbox{and} \\
\rho_{\mbox{ns}} 
& \approx & 10^{17} \mbox{kg}/\mbox{m}^3 \approx 1.4 \times 10^{-4} /M_\odot^2 
\eea
are the mean densities of our sun, a typical white dwarf, a typical neutron star respectively. 

Finally for simplicity and consistency  we will typically study black holes that either start or finish with solar mass. Such black holes are not fully realistic since astrophysical black holes forming from stellar collapse have masses of $4-15 M_\odot$. However the conversion of our examples to 
more realistic masses is easy. For example to retool our examples with $6 M_\odot$ as the standard mass, simply multiply the given times and lengths by 6 and divide any densities by 36. 

\section{Slowly evolving Vaidya horizons}
\label{vaidya}

\subsection{General considerations}

The Vaidya solution is the simplest example of a dynamical black hole spacetime \cite{eric}. 
It models the collapse of null dust and is described by the metric
\bea
ds^2 &=& - \left(1- \frac{2m(v)}{R} \right) dv^2 + 2 dv dR + R^2 d \Omega^2 \, , 
\eea
where $v$ is an advanced time coordinate, $m(v)$ measures the mass of the black hole on a hypersurface of constant  $v$ and the infalling null dust has stress-energy tensor  
\bea
T_{ab} &=& \frac{dm/dv}{4 \pi R^2} [dv]_a [dv]_b \, .
\eea

We consider spherically symmetric MTTs.  Scaling the null vectors as 
\bea
\ell^a = \left[1, \frac{1}{2} \left(1 - \frac{2m(v)}{R}\right) , 0 , 0  \right] \, \, \,  \mbox{and} \, \, \,  n^a = [0, -1, 0, 0] \, , 
\eea
it is a quick calculation to show that 
\bea
\tl = \frac{1}{r} \left( 1 - \frac{2m(v)}{R} \right) \, \, \, \mbox{and} \, \, \, \tn = - \frac{2}{R} \, 
\eea
and so there is a FOTH located at $R_H = 2m(v)$. Then from 
$\Lie_{\mathcal{V}} \tl = 0$ (since $\tl = 0$ everywhere on an MTT), 
a straightforward calculation shows that
\bea
C = \frac{dR_H}{dv} =  2 \frac{dm}{dv} \, 
\eea
and so the slowly evolving parameter is
\bea
\epsilon^2 \equiv C \tn^2 (R_H)^2 = 8 \frac{dm}{dv} \, .
\eea
Then, conditions S\ref{s1} and S\ref{s2} will both be satisfied if $dm/dv$ is sufficiently small. 
Given the spherical symmetry, S\ref{s3} and S\ref{s4} 
are also easily checked and seen to hold in this case. 
Thus for Vaidya we have a slowly evolving horizon whenever $dm/dv$ is sufficiently small. 

Further $T_{ab} \ell^a n^b = T_{ab} n^a n^b = 0$ and so from (\ref{kap0}) it is clear that 
\be
\kappa_{\cV} = \frac{1}{4 m(v)} 
\ee
exactly. 
Then around a given value $v_o$ we can Taylor expand to find : 
\bea
\kappa_{\cV} = \frac{1}{4 m(v_o)} \left(1 -  \frac{\epsilon^2}{8 m(v_o)} (\Delta v)^2 
+ \mbox{O} (\epsilon^4) \right) \, . 
 \eea
%
This  is in agreement with (\ref{zeroth}) (with $\kappa_1 = 0$).  The surface gravity is constant over a given MTS and further only changes slowly between MTSs. 

In this case the first law actually holds exactly for \emph{any} value of $dm/dv$. Specifically, both sides of equation (\ref{firstLaw}) evaluate exactly as $8 \pi dm/dv$. As we shall soon see however, this is a special property of Vaidya. In most cases this form of the first law only holds for slowly evolving horizons.  

\subsection{Piecewise linear example}

Let us now restrict to a specific example to get a physical feeling for when Vaidya is slowly evolving. 
Consider a piecewise linear mass function : 
\bea
m(v) = \left\{
\begin{array}{ll}
1 & v \leq 0 \\ 
1+ \alpha v & 0 \leq v \leq v_o\\
1+ \alpha v_o & v>v_o
\end{array} \right.  \, .
\eea
Thus, a black hole is irradiated with null dust for a finite period of time. For definiteness we assume the solar unit system discussed above and so this corresponds to a solar mass black hole that evolves to a
$(1+\alpha v_o)$ solar mass black hole. 

Since there is no natural way to 
measure time along a spacelike horizon, a 
common problem in all of these examples is to decide on a reasonable reference frame in which to 
judge whether we would expect the horizon to be slowly evolving . In this case, we make use of a fleet of observers who use rockets to hold themselves at constant areal radius $R_o$. Then, these observers will see a total mass of $\alpha v_o$ fall past them during time 
\be
T =  \int_0^{v_o} \sqrt{1 - \frac{2m(v)}{R_o}} dv <  \alpha v_o \, , 
\ee 
with the inequality coming close to saturation for very large $R_o$. 

Now, though these observers can't actually see the evolving horizon, let us consider the situation for  
$\alpha = 1/80000$ and $v_o \approx 2.1 \times 10^5 M_\odot$ (one second in standard
units). Then, our observers would see $\alpha v_o \approx 2.6$ solar masses of material sweep past 
them in less then a second. However, at the horizon a quick calculation shows that 
$\epsilon^2  = 8 \alpha = 10^{-4}$. Thus, even in this fairly dramatic situation the horizon would 
be slowly evolving to order $\epsilon \approx 0.01$. 

\section{Slowly evolving Tolman-Bondi horizons}
\label{TB}

\subsection{General considerations}

Our next series of examples are Tolman-Bondi spacetimes and so consist of spherically symmetric 
clouds of timelike dust collapsing to form (or falling into) a black hole. These may be 
described in the following way (a more complete discussion of these spacetimes and their properties with particular emphasis on marginally trapped tube evolutions may be found in \cite{mttpaper}). 

We will consider gravitationally bound Tolman-Bondi spacetimes which contain an instant
of time symmetry. That is, they contain a spacelike three-surface $\Sigma_o$ with intrinsic metric
\bea
ds^2 = \frac{dr^2}{1 - 2 m(r)/r} + r^2 (d \theta^2 + \sin^2 \theta d \phi^2) 
\eea
and extrinsic curvature $K_{ab} = 0$. $m(r)$ is a mass function defined by 
\be
m(r) = \int_o^r d\tilde{r} (4 \pi \tilde{r}^2) \rho_o(\tilde{r})\, ,  \label{mass}
\ee
where $\rho_o$ specifies the initial distribution of (stationary) dust on $\Sigma_o$.  Thus on this initial 
surface the dust has stress-energy tensor 
\be
T_{ab} = \rho_o \hu_a \hu_b \, , \label{TBT}
\ee
 where $\hu^a$ is the unit timelike normal to the surface. Note too that the spacelike assumption means that $r > 2 m(r)$ everywhere and there are no black holes in this initial data.

Evolving this data forward in time, the dust falls inwards and one obtains the full four-dimensional metric 
\be
ds^2 = - d\tau^2 + \frac{(R'(\tau,r))^2}{1 - 2m(r)/r}dr^2 
+ R^2(\tau,r)\, d \Omega^2 \, , 
\ee
where $\tau$ records proper time as measured by observers comoving with the dust. 
$R(\tau,r)$ is the areal radius at time $\tau$ of the spherical set of observers who had initial areal radius
$r$ and must be a solution of 
\be
\dot{R}(\tau,r) \equiv \frac{\partial R(\tau,r)}{\partial \tau} = 
- \sqrt{ \frac{2m(r)}{R(\tau,r)}}   \sqrt{1 - \frac{R(\tau,r)}{r} }\, . 
\label{EErem}
\ee
$R'(\tau,r) = \partial R(\tau,r) / \partial r$ and the stress-energy tensor keeps 
the form $T_{ab} = \rho(\tau,r) \nabla_a \tau \nabla_b \tau$.

In fact, this equation has an exact parametric solution given by:  
\begin{eqnarray}
R(\eta,r) &=& r \cos^2\left(\frac{\eta}{2}\right) \, \, \,  \mbox{and} \label{Reta} \\
\tau(\eta,r) &=&  \left(\frac{r^3}{8m(r)}\right)^{1/2} \mspace{-5mu}(\eta 
+ \sin \eta)  \label{teta} 
\end{eqnarray}
for a parameter $\eta \in [0 , \pi]$.  In our examples we will make use of this solution to 
generate the full spacetime evolutions. However, as an immediate application we note that it immediately follows that shells with initial areal radius $r = r_o$ will collapse to zero area at
\be
\tau_c (r_o) = \pi \sqrt{\frac{r_o^3}{8m(r_o)}} \, . 
\ee
For all of the examples which we consider, this will be a strictly increasing function. In cases where it isn't, shell-crossing singularities occur (see for example \cite{gonc}). 

For our analysis we need a pair of cross-normalized null vectors which
we pick to have the form : 
\bea
\ell_a &=& - \sqrt{1- 2m(r)/r} [d \tau]_a + R'(\tau,r)[dr]_a 
\, \, \,  \mbox{ and} \\
n_a &=&  - \frac{1}{2\sqrt{1- 2m(r)/r} } [d \tau]_a - \frac{R'(\tau,r)}{2 (1-2m(r)/r)} \nn
[dr]_a \, . 
\eea
Then, it is a straightforward calculation to show that
\be
\tl = \frac{2 (1-2m(r)/r)}{R(\tau,r)} \left( \frac{\dot{R}(\tau,r)}{\sqrt{1-2m(r)/r}} + 1\right)  \label{Xtl}
\ee
and
\be
\tn = \frac{1}{R(\tau,r)} \left(\frac{\dot{R}(\tau,r)}{\sqrt{1-2m(r)/r}} - 1 \right)  \, . 
\ee
Combining the expression for $\tl$  with (\ref{EErem}) it is easy to see that the trapping horizons may 
be found whenever $R(\tau,r) = R_H =  2 m(r)$ and on those horizons 
\bea
\left. \tn \right|_{H} = - \frac{2}{R_H} < 0 \, . 
\eea
Thus the surface $R(\tau,r) = 2m(r)$ is a FOTH.

With this choice of scaling for the null normals we find that 
\bea
C = 2 \left( 1 - \frac{2m(r)}{r} \right) \frac{m'(r)}{R'(\tau, r) - m'(r)} 
\eea
and so the first two slowly evolving conditions (S\ref{s1} and S\ref{s2}) will be met if
\bea
\epsilon^2 & \equiv & C \tn^2 R_H^2  
  =  \frac{8 m'(r)}{R'(\tau,r)-m'(r)}     \label{SEP}
\eea
is sufficiently small.

We have not given an explicit form for the parameter $v$. However, note that
\bea
\Lie_{\cV} \tau = 1 - \frac{C}{2(1 - 2 m(r)/r)} 
\eea
and so when $r \gg 2 m(r)$ (which it usually will be in our examples) and the horizon is slowly 
evolving, $\Lie_{\cV} \tau \approx 1$.  
The final two conditions (S\ref{s3} and S\ref{s4})
are also easily shown to hold if $\epsilon$ is sufficiently small. 

Next, consider the surface gravity. From (\ref{kap0}) and (\ref{TBT}) we have
\bea
\kappa_{\cV} =  \frac{1}{2R_H} \left( 1 + \frac{C}{2 ( 1 - 2m(r)/r) } \right) \, . 
\eea
Thus, when $r \gg 2 m(r)$, $\kappa_{\cV}$ will be $1/2R_H$ to order $\epsilon^2$ and it is also clear
that $\dot{\kappa}_{\cV} \approx -C/(2R_H^2)$ is small. 

Then, if as usual we assume that $r \gg 2m(r)$ we have 
\bea
\frac{dm}{dv} \approx\frac{\kappa_{\cV} }{8 \pi G} \frac{da_H}{dv} 
\eea
and 
\bea
\int_{H_v} \mspace{-12mu} \sqrt{\tq} T_{a b} \ell^a \ell^b = \frac{C}{2 + C (1 - 2m(r)/r)} \approx \frac{dm}{dv} \, ,
\eea
and so the first law holds to second order as expected. 


\subsection*{Accretion}

For our first set of explicit examples, we study accretion of a (spherical) dust shell of timelike dust 
onto an existing (and initially isolated) horizon \cite{mttpaper}. To model this we start with both a Schwarzschild solution 
of mass $m_o$ and the Tolman-Bondi solution which evolves from a dust density distribution of 
\bea
\rho_o (r) = \frac{\mu e^{- (r/r_o-\alpha)^2}}{2 r_o^3 \pi^{3/2} (1+2 \alpha^2)} \frac{1}{M_\odot^2}
\eea
on the initial surface $\Sigma_o$. Here, $\mu$ is the total shell mass\footnote{The total mass will actually be very slightly smaller than $\mu$ due to the excision of the region inside $r=2m_o$ however for this example the difference will be negligible.}, $\alpha$ locates its centre in units of $r_o$,
and $r_o$ characterizes its thickness : $86.5\%$ of the mass lies within $\pm r_o$ of $\alpha r_o$, 
and $99.4 \%$ lies within $\pm 2 r_o$. 

Our desired spacetime is constructed from these two with the help of a bit of spacetime surgery.
We remove the region inside $r=2m_o$ in the Tolman-Bondi spacetime and replace it with the corresponding part of Schwarzschild. This can be done smoothly and gives a spacetime in which the dust shell collapses onto the black hole. The mass function of $\Sigma_o$ is then: 
\bea
m(r) = m_o + 4 \pi \int_{2M_o}^r \mspace{-16mu} d \tilde{r} \{\tilde{r}^2 \rho_o (\tilde{r}) \}
\eea

For definiteness fix the parameters of this spacetime as follows. Working in solar mass units, 
take $m_o = 1$ so that the initial black hole has the same mass as our sun. Next, fix
$\alpha = 4$ and $r_o = 500$ so that the dust shell has peak density at 
$\alpha r_o = 2000 R_\odot \approx 2800 \mbox{km}$ and almost all of the dust ($99.4 \%$)
 lies between $2000 R_\odot \approx 1400 \mbox{km}$ and $3000 R_\odot \approx 4200 \mbox{km}$. 
Further, with these choices of $\alpha$ and $r_o$, $\rho_o \approx 2 \times 10^{-11} \mu$ and so for
$\mu \sim 1$ peak density is about that of a white dwarf. 

\begin{figure}
\resizebox{9cm}{!}{\includegraphics{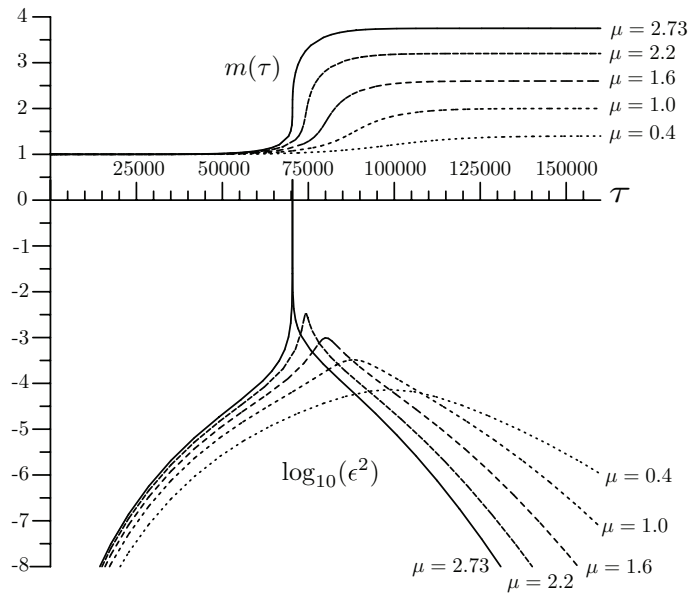}}
\caption{The evolution of an initially solar mass black hole 
as dust shells of various masses collapse onto it. 
The bottom part of this graph shows the corresponding values of the slowly evolving parameter. It is clear that while all of the MTTs start and finish as slowly evolving, for larger mass shells they are not 
always slowly evolving throughout their evolution.}
\label{accrete}
\end{figure}

Then, with the help of the parametric solution given in (\ref{teta}) and (\ref{Reta}) we can study the full
spacetime. Specifically for $\mu = 0.4$, $1.0$, $1.6$, $2.2$, and $2.73 M_\odot$ the evolution of the horizon
is shown in Figure \ref{accrete}. The top half displays its mass as a function of $\tau$
while the bottom part shows the corresponding values of the slowly evolving parameter. In interpreting
this graph, keep in mind that since $\tau$ is given in solar mass units, the time span shown corresponds
to about one second.  

These values of $\mu$ were deliberately chosen to be on the boundary between slowly and non-slowly
evolving horizons. It is clear that each horizon will be slowly evolving at the beginning and
end of the accretion, this will usually not be the case during the peak inflow of matter. Note however
that for $\mu = 0.4$, $\epsilon < 0.01$ throughout its evolution and so this case could reasonably be 
considered as slowly evolving throughout. This is true
despite the fact that observers comoving with 
the dust would experience the $0.4 M_\odot$ of dust falling into the black hole in less than one second. 

\subsection*{Collapse}
For our next example we consider what would seem to be an even more dramatic situation --- the 
gravitational collapse of a dust cloud to form a new black hole. Again we work with a Tolman-Bondi solution this time choosing 
\be
\rho_o = \frac{\mu}{\pi^{3/2} r_o^3} e^{-(r/r_o)^2} \left(\frac{1}{M_\odot^2} \right)\, , \label{DensCollapse}
\ee
where $\mu$ is the total mass of the dust cloud, $r_o$ and $r_o$ quantifies its width : in this case
$42.8 \%$ of the mass lies within $r_o$ of the centre, $95.4\%$ within $2r_o$ and 
$99.96 \%$ within $3r_0$. 

To be concrete we pick $r_0 = 1000 M_\odot \approx 1400 \mbox{km}$ and so this is a (spatially) 
small dust cloud. Further, with this choice the central density of the initially stationary dust is 
$\rho(0) \approx 1.8 \times 10^{-10} \mu$ and so is somewhat higher than the average density of a 
white dwarf. We explicitly consider the cases $\mu =0.25, 0.5, 1.0, 2.0$, and $4.0 M_\odot$. 

\begin{figure}
\resizebox{9cm}{!}{\includegraphics{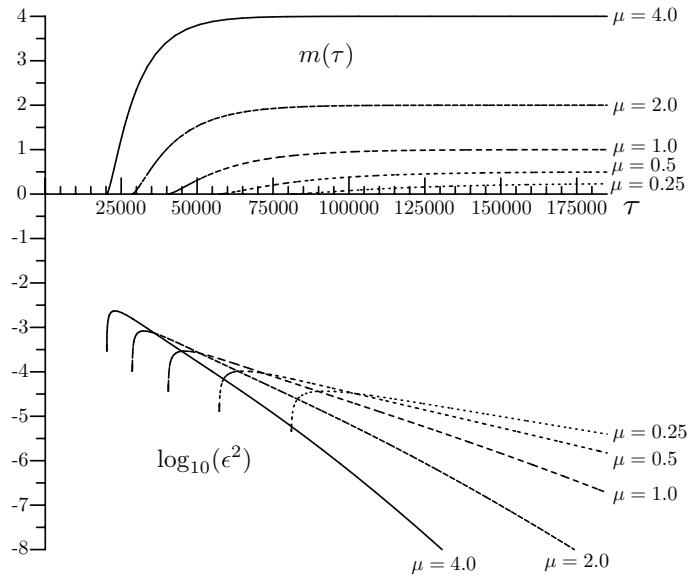}}
\caption{The collapse of dust clouds of various initial masses to form a black hole.
The bottom part of this graph shows the corresponding values of the slowly evolving parameter. While
in all of the cases the MTTs asymptote into SEHs, note that for $\mu = 0.25$ the horizon is slowly evolving throughout its formation and evolution. }
\label{collapse}
\end{figure}

Then, the evolution as a function of $\tau$ is shown in Figure \ref{collapse}. From the top half 
of that figure we see that horizons form between $\tau \approx 20000 $ and $\tau \approx 85000$ (that is from $0.1$ to $0.4$ seconds after the collapse begins) and in all cases 
everything is essentially over by $\tau =175000$ (before a full second is over). Turning to the bottom
half of the graph we then see that while all of the horizons asymptote into the slowly evolving regime
in the expected way, the higher mass solutions begin outside of this regime. Meanwhile, 
for $\mu = 0.25 M_\odot$, $\epsilon < 0.01$ throughout the entire collapse. Thus we see that under 
the correct circumstances horizon formation can actually be a slowly evolving process. 

Note too that again we have choosen our examples to be on the borderline between slowly and 
non-slowly evolving horizons. Lower mass and/or more dispersed clouds all fall even more firmly into the slowly evolving regime while more massive and/or more concentrated clouds will have larger
values of $\epsilon$.

\section{Tidal Examples}
\label{tidal}

Finally, we (perturbatively) break the spherical symmetry of the previous examples and consider 
the evolution of a black hole sitting in a slowly changing external field. Physically, such a situation 
would occur  as a black hole of mass $M$ moves through an ``external" gravitational 
field whose characteristic radius of curvature $\mathcal{R}$ as well as characteristic rates of change in 
time $\mathcal{T}$ and space $\mathcal{L}$ are all much larger than $M$. In certain regimes, 
one can split the total gravitational field into an external field and that part arising 
from the black 
hole and its motion. In particular, close to the hole one can describe the total field as that of the hole 
plus a perturbation from the external field while far from the hole the roles reverse. A specific example
of such a configuration of fields would be
a small black hole moving in a slow orbit around a much larger one. 
The formalism necessary to describe this situation was first developed over 20 years ago 
\cite{th, zhang} though we shall most closely follow the recent work of Poisson
\cite{ericBig, ericPRL} which pushes this formalism to the higher order that is necessary for 
our calculations. 

\subsection{The Set-up}
In a little more detail, we begin with the external spacetime and the timelike geodesic $\gamma$ 
that the black hole will ultimately follow. Then in a comoving coordinate system the metric in some neighbourhood of the geodesic may be written as a perturbation of Minkowski space. In particular, at the origin of that system, coordinate time will coincide with proper time along the curve. The coordinate
system will be based on the timelike unit vector $u^a$ up the geodesic plus a orthonormal 
spacelike triad $e^a_i$, $i \in [1,2,3]$,  which is parallel propagated along $\gamma$. 

We assume that the external spacetime is vacuum and so the spacetime curvature 
is described entirely by the Weyl tensor $C_{abcd}$. With the help of the completely anti-symmetric 
Levi-Cevita tensor, this may be decomposed into two symmetric, trace-free, 
three-dimensional tensors:
\be
\mathcal{E}_{ij} = e^a_i e^b_j C_{acbd} u^c u^d \, \, \,    \mbox{and} \, \, \,  
\mathcal{B}_{ij} = \frac{1}{2} e^a_i e^b_j \varepsilon_{a c}^{\; \; \; \; d e} C_{debf} u^c u^f \, , \label{CandE}
\ee
which are respectively referred to as the ``electric" and ``magnetic" Weyl tensors. The perturbation
expansion of the metric is then constructed from these tensors and their derivatives. 

To do this, we begin by making our earlier assumption about the magnitudes of quantities
precise in the following way. 
First, the assumption about the radius of curvature of the spacetime says that the components 
of the electric and magnetic Weyl tensors 
are on the order of (or smaller than) $\mathcal{R}^{-2}$. Next, with 
respect to geodesic time we have both $\dot{\mathcal{E}}_{ij}$ and $\dot{\mathcal{B}}_{ij} \lesssim \mathcal{R}^{-2} \mathcal{T}^{-1}$. Similarly the homogeneity assumption says that on extending
the tetrad into a neighbourhood of $\gamma$, spatial derivatives of these quantities such as $\Lie_{e_i} \mathcal{E}_{ij} \lesssim \mathcal{R}^{-2} \mathcal{L}^{-1}$. For notational convenience we reduce 
these three parameters to one, defining $\mathcal{X} = \mbox{min} \{ \mathcal{R}, \mathcal{L}, \mathcal{T} \}$ and consider the expansion in powers of $\mathcal{X}^{-1}$. Thus, the components of the 
electric and magnetic tensors are $\lesssim \mathcal{X}^{-2}$ while the derivatives of these quantities are $\lesssim \mathcal{X}^{-3}$. 

Now, in Fermi normal coordinates based on the freely falling $\{u^a, e^a_i\}$ tetrad the metric may be
expanded in the following way. First, since the tetrad is orthonormal, the metric is Minkowski to 
zeroth order. Next at dipole ($\mathcal{X}^{-1}$) order there are no corrections since the 
frame is freely falling. The first corrections occur at quadropole ($\mathcal{X}^{-2}$) order where 
they are are constructed from the $\mathcal{E}_{ij}$ and $\mathcal{B}_{ij}$. Terms at octopole
($\mathcal{X}^{-3}$) order depend on the temporal (along $\gamma$) and spatial 
(in the $e^a_i$ directions) derivatives of these quantities. Higher order terms similarly depend on 
higher order derivatives though they will not concern us in this paper.
 
%

If we insert a mass $M \ll \mathcal{X}$ Schwarzschild black hole into the spacetime and set it to travel along the worldline, we can characterize the induced perturbations near the hole 
($2M <  r \ll \mathcal{X}$) in terms of those same quantities. Specifically in ingoing Eddington-Finkelstein type coordinates \cite{ericPRL}, 
%
%
\bea
ds^2 &=& -\left(1 - \frac{2M}{r} - \delta g_{vv} \right) dv^2 - 2 dv dr + 2 \delta g_{vA} dv d\theta^A
+\left( r^2 \Omega_{AB} + \delta g_{AB} \right) d\theta^A d \theta^B \, , \label{tidalMetric}
\eea
where $\Omega_{AB} = \mbox{diag} [1,  \sin^2 \theta]$, and  
$A,B \in \{1,2\}$ with $\theta^1 = \theta$ and $\theta^2 = \phi$. The tidally induced perturbations then 
take the form: 
\bea
\delta g_{vv} & = &  r^2 e_1 \mathcal{E}^{\mathsf{q}} 
- \frac{r^3}{3} \left(e_2 \dot{\mathcal{E}}^{\mathsf{q}}
- e_3    {\mathcal{E}}^{\mathsf{o}} \right)  + \mbox{O}(\mathcal{X}^{-4})  \label{MetComp} \\
\delta g_{vA} & = &  -\frac{2r^3}{3} 
\left(e_4 \mathcal{E}^{\mathsf{q}}_A - b_4 \mathcal{B}^{\mathsf{q}}_A \right) 
+ \frac{r^4}{3} 
\left(e_5 \dot{\mathcal{E}}^{\mathsf{q}}_A - b_5 \dot{\mathcal{B}}^{\mathsf{q}}_A \right) 
- \frac{r^4}{4} 
(e_6 \mathcal{E}^{\mathsf{o}}_A - b_6 \mathcal{B}^{\mathsf{o}}_A )+ \mbox{O}(\mathcal{X}^{-4})  \nn \\
\delta g_{AB} & = & 
- \frac{r^4}{3} 
(e_7 \mathcal{E}^{\mathsf{q}}_{AB} - b_7 \mathcal{B}^{\mathsf{q}}_{AB} )
- \frac{5r^5}{18} 
(e_8 \dot{\mathcal{E}}^{\mathsf{q}}_{AB} - b_8 \dot{\mathcal{B}}^{\mathsf{q}}_{AB} ) 
- \frac{r^5}{6} 
(e_9 \mathcal{E}^{\mathsf{o}}_{AB} - b_9 \mathcal{B}^{\mathsf{o}}_{AB} ) + \mbox{O}(\mathcal{X}^{-4})  \nn \, ,
\eea
where the exact definitions of the various $\mathcal{E}$ and $\mathcal{B}$ terms are given in appendix 
\ref{irtf}. The $\mathsf{q}$ and $\mathsf{o}$ superscripts indicate the multipole classification (and hence expansion order)
of the various terms while the over-dots indicate derivatives with respect to $v$ 
(and so lower the order of dotted terms by a factor of $\mathcal{X}$). 
The full expressions for the $e_n$ and $b_n$ are given in appendix \ref{appB}, though here we note 
that if $M=0$ (ie. no hole) the $e_n$ and $b_n$ are all unity while if $M \neq 0$ they are functions 
of $r$ alone. These functions go to unity as $r \rightarrow \infty$ and vanish at $r=2M$
except for $e_7 (2M) = 1/2$, $e_9(2M) = 1/10$, $b_7(2M) = -1/2$ and $b_9(2M) = -1/10$
(though the derivatives are, in general, non-zero). 

\subsection{Locating a horizon}

In general, locating an MTS is not a trivial task and indeed after a glance at
the metric (\ref{tidalMetric}), one might suspect that that will be the case here. 
In fact, such worries are unfounded. It turns out that to the order in which we are working, 
 $r = 2M$ foliated by $v = \mbox{constant}$ surfaces is a FOTH. 

The see this, first note that the induced metric on a $r=2M$ and $v=\mbox{constant}$ is 
\bea
\tq_{AB} & = & 4M^2 \left( \Omega_{AB} - \frac{2}{3} M (\mathcal{E}^{\mathsf{q}}_{AB} 
+  \mathcal{B}^{\mathsf{q}}_{AB}) 
- \frac{2}{5} M^2 (\mathcal{E}^{\mathsf{o}}_{AB} +  \mathcal{B}^{\mathsf{o}}_{AB} ) \right)
+ \mbox{O}(\mathcal{X}^{-4}) \label{tq}
\eea
and so the corresponding area element is
\bea
\sqrt{\tq} = 4 M^2 \sin \theta + \mbox{O}(\mathcal{X}^{-4}) \, ,  \label{TidalArea}
\eea
which follows from the fact that the perturbations are tracefree 
with respect to the unperturbed (spherical) metric.  
Next, up to rescalings, the future outward pointing null normal vector field to these surfaces takes the simple
form  
\bea
\ell_a  =  [dr]_a & \Leftrightarrow & \ell^a =  \left[ \frac{\partial}{\partial v} \right]^a 
+ \mbox{O}(\mathcal{X}^{-4}) \,  \label{ell}
\eea
thanks to the vanishing of the majority of the $e_n$ and $b_n$ at $r=2M$. 
Then 
\bea
\tl = \frac{1}{\sqrt{\tq}} \Lie_\ell \sqrt{\tq} &=& \frac{1}{\sqrt{\tq}} \frac{d }{dv} \sqrt{\tq}
+ \mbox{O}(\mathcal{X}^{-4}) \label{tl} \\
&=&  0 +  \mbox{O}(\mathcal{X}^{-4}) \nn
\eea
by direct calculation. 




Shifting our attention to the inward null expansion, 
the future inward pointing null normal vector field to the two-surfaces (cross-normalized
so that $\ell \cdot n = -1$) is  
\bea
 n_a  =  -[dv]_a & \Leftrightarrow & n^a = - \left[ \frac{\partial}{\partial r} \right]^a 
 + \mbox{O}(\mathcal{X}^{-4}) \, ,  \nn
\eea
and so 
\bea
\tn & = & \tq^{ab} \nabla_a n_b = -\frac{1}{M} +  \mbox{O}(\mathcal{X}^{-4})  \, . \label{tn}
\eea

Finally another direct calculation shows that 
\bea
\delta_n \tl & = & - \frac{d}{dr} \tl + \mbox{O}(\mathcal{X}^{-4}) \\
&  \approx & - \frac{1}{4 M^2} + \mbox{O}(\mathcal{X}^{-2}) . \nn
\eea
The ``$\approx$" in the last line indicates that the metric can be used to calculate this quantity to higher 
order (actually $\mathcal{X}^{-4}$) than shown, but this has not been done since all that we needed to 
show was that $\delta_n \tl < 0$. 

Thus, combining the results of these three calculations we see that the $r=2M$ surface foliated by $v = \mbox{constant}$ surfaces is a FOTH as asserted. 

\subsection{Horizon properties}

Next, it is straightforward to see that this FOTH is actually a slowly evolving horizon. First from 
equations (\ref{Adot}) and (\ref{TidalArea})
it is immediate that the expansion parameter $C$ is at most of order $\mathcal{X}^{-4}$. Combining
this with the expression (\ref{tn}) for $\tn$ it is easy to see that conditions S\ref{s1} and S\ref{s2}
hold. 

Checking the other slowly evolving conditions requires the calculation 
of a few more quantities. Since the spacetime is vacuum, the matter terms clearly vanish 
while the other quantities are:
\bea
 \sigma^{(n)}_{AB} & = &  \frac{8}{3} M^3 \left( \mathcal{E}^{\mathsf{q}}_{AB} 
 + \mathcal{B}^{\mathsf{q}}_{AB} \right) 
 + 8 M^4 \left( - \frac{11}{9} \dot{\mathcal{E}}^{\mathsf{q}}_{AB} 
 +  \dot{\mathcal{B}}^{\mathsf{q}}_{AB}\right) + \mbox{O} (\mathcal{X}^{-4}) \,  ,  \\
\tom_{A} & = & - \frac{4}{3} M^2 \left( \mathcal{E}^{\mathsf{q}}_{A} 
- \mathcal{B}^{\mathsf{q}}_{A} \right) 
- \frac{1}{9} M^3 
\left(16 \dot{\mathcal{E}}^{\mathsf{q}}_{A} +  3 \mathcal{E}^{\mathsf{o}}_{A} 
- 3 \mathcal{B}^{\mathsf{o}}_{A}\right)  + \mbox{O}(\mathcal{X}^{-4}) \, \, \, \mbox{and}  \\
\tilde{\mathcal{R}} &  \approx &  \frac{1}{4M^2} + \mbox{O}(\mathcal{X}^{-1}) \, ,  
\eea
where the ``$\approx$" in the expression for the Ricci scalar again indicates that the metric actually defines
this quantity to higher order than indicated. Then it is easy to see that conditions S\ref{s3} and S\ref{s4} also
hold and so these tidally distorted black holes have a slowly evolving horizon of order 
$\epsilon = \mathcal{X}^{-2}$ at $r=2M$.  Further, though we will not show the explicit calculations here, it is not hard to show that the standard two-sphere rotational Killing vectors remain as approximate symmetries (as defined in \cite{prl}) of the perturbed horizon. 

We can then consider the mechanics of this horizon. First
\bea
\kappa_{\cV} &=& \frac{1}{4M} \left( 1  + \frac{16}{3} M^3 \dot{\mathcal{E}}^{\mathsf{q}} \right) 
+ \mbox{O}(\mathcal{X}^{-4} )\, .
\eea
Thus, the surface gravity is approximately constant over each $H_v$ and further only changes 
slowly in time as would be expected for a quasi-equilibrium state. Note too that 
the expression for the surface gravity matches that found in \cite{ericPRL}  (though this isn't
especially surprising for this spacetime since the event horizon overlaps this FOTH
to the order in which we are working). 

The first law also holds. Since $C$ is at most of order 
$\epsilon^2 = \mathcal{X}^{-4}$, the left hand side of equation (\ref{firstLaw}) vanishes to this order. 
To evaluate the right-hand side we need an expression for $\sigma^{(\ell)}_{AB}$ 
which we calculate as
\bea
\sigma^{(\ell)}_{AB} = \frac{1}{2} \Lie_\mathcal{V} \tq_{AB} 
& = & - \frac{4}{3} M^4 \left( \dot{\mathcal{E}}^{\mathsf{q}}_{AB} 
+ \dot{\mathcal{B}}^{\mathsf{q}}_{AB} \right) \label{sigell} 
- \frac{4}{15}  M^5 \left( \dot{\mathcal{E}}^{\mathsf{o}}_{AB} + \dot{\mathcal{B}}^{\mathsf{o}}_{AB} \right) 
+  \mbox{O} (\mathcal{X}^{-5}) \, . 
\eea
As in the expansion calculation (\ref{tl}), the Lie derivative is simply evaluated as the time derivative with 
respect to $v$. Then given the form of (\ref{tq}) and the assumption that taking derivatives with respect to 
$v$ lowers the order of metric quantities by a factor of $\mathcal{X}$, the above expression follows 
directly.\footnote{While we have only claimed fourth order accuracy in (\ref{ell}), in fact
horizon-locking coordinate choices have been made the mean that this shear may quantity may 
be calculated directly from the metric  to this higher order. 
This is confirmed by a gauge-invariant calculation based on the 
even-parity and odd-parity master functions \cite{ericPRL, ericCom, karleric}. } 
Thus, the right-hand side of that equation evaluates as:
\bea
\frac{1}{8 \pi G} \int_{H_v} d^2 x \sqrt{\tq}  \sigma^{(\ell)}_{ab} \sigma^{(\ell) ab}
= \frac{16}{45} M^6 \left(\dot{\mathcal{E}}_{ij} \dot{\mathcal{E}} ^{ij} 
+ \dot{\mathcal{B}}_{ij} \dot{\mathcal{B}} ^{ij}\right) 
+ \frac{16}{4725} M^8 \left( \dot{\mathcal{E}}_{ijk} \dot{\mathcal{E}} ^{ijk} 
+ \dot{\mathcal{B}}_{ijk} \dot{\mathcal{B}} ^{ijk}\right) + \mbox{O}(\mathcal{X}^{-9})  \, ,  \label{RHS} 
\eea
where the computation makes use of equations 
(\ref{TidalArea}),  (\ref{sigell}), and (\ref{intFirst}-\ref{intLast}) and is straightforward but quite lengthy. 
Since both of the left and right hand sides of (\ref{firstLaw}) are zero to order $\mathcal{X}^{-3}$ the first law holds.

More interesting though is the result (\ref{RHS}). As it is the ``square" of a quantity known accurately 
to order $\mathcal{X}^{-4}$ 
we can trust it to order $\mathcal{X}^{-8}$.  
Thus, we see that in this case the first law allows us to calculate the rate of change of the mass
even though we cannot see those changes directly in the approximation that we are using. 
Note too that this result matches those found in \cite{ericBig, ericPRL} though with the difference 
that our calculation gives a snapshot rather than time averaged value for $dm/dv$. 

Finally we again consider the physical situation under which this is a slowly evolving horizon. Taking
$M = M_\odot$ the time scale of the changing fields must be significantly larger than 
$T_\odot$. Taking $\mathcal{T} = 100 T_\odot$ and assuming that $\mathcal{X} \sim \mathcal{T}$
this gives $\frac{dM}{dv} \sim 10^{-12} \sim 10^{-7} M_\odot/\mbox{sec}$.

\section{Conclusion}
This paper provides concrete examples of slowly evolving horizons and demonstrates that they
correspond to situations that one would intuitively consider to be slowly evolving. Thus we saw that
Vaidya and Tolman-Bondi black holes with sufficiently small mass flows are slowly evolving as is 
the tidally perturbed Schwarzschild black hole. In the Tolman-Bondi examples we studied
horizons that start out isolated, transition through the slowly evolving regime as an infalling dust cloud 
begins to accrete, pass out of it as the bulk of the matter hits, and then return to that regime 
as the last dregs of dust fall in and the horizon asymptotes towards isolation. 
Similarly, in the aftermath 
of a black hole formation from a dust cloud collapse, the MTT is slowly evolving as it approaches isolation. 
A bit more suprisingly, we saw that in some cases the formation of a black hole
during gravitational collapse can be characterized by an SEH throughout its entire evolution. 

It also became clear that, at least in the spherically symmetric case, 
we need to pay careful attention to mass scales in intuitive 
judgements as to whether a particular horizon is slowly evolving or not. 
For approximately solar mass black holes 
accretion has to be truly spectacular in order to have a non-SEH MTT. By contrast, for a supermassive
$10^9 M_\odot$ hole the characteristic time scale would be over an hour and the characteristic density 
would be $\sim 1000 \mbox{kg}/ \mbox{m}^3$. 
Thus, in contrast to smaller black holes where matter of density somewhere between that 
of a white dwarf and a neutron star is needed to move an MTT out of the slowly evolving regime, 
here only densities about that of liquid water are needed. 

Given the orders of magnitude involved in these examples, it seems like that 
the intuition that we have gained here may apply, at least qualitatively, even away from 
spherical symmetry. Thus, the astrophysics of stellar mass black holes is probably largely ``slow" (apart from 
especially dramatic events like black hole mergers, neutron star/white dwarf captures, or some formations)
while that of super-massive black holes SEHs will often not be in that regime. For example, one suspects 
that the capture of even a small star by a super-massive black hole might break the SEH conditions
(and it might even cause a horizon ``jump" as discussed in \cite{mttpaper,SKB}). We hope to see this
conjecture tested by future numerical simulations of more realistic astrophysical situations. 

In developing any theory it is very useful to have specific, easily handled, examples in mind against which one can test one's ideas. This paper provides some of these examples for slowly evolving horizons and so helps to form both our mathematical and physical intuition about the quasi-equilibrium states of black holes. 

\section*{Acknowledgements} The authors would like to thank the participants of Black Holes V (Banff Centre, 2005) and CCGRRA 11 (University of British Columbia, 2005), and in particular Eric Poisson, for useful comments and discussions on this work. They also thank Stephen Fairhurst for his contributions and the 
referee for many useful suggestions.  
Financial support was provided by the Natural Sciences and Engineering Research Council of Canada. 

\begin{appendix}

\section{Irreducible Tidal Fields} 
\label{irtf}

To octopole order, the terms appearing in the metric are \cite{ericPRL}:
\be
\begin{array}{lll}
\mathcal{E}^{\mathsf{q}}&=& \mathcal{E}_{ij} \Omega^i \Omega^j  \, , \\
\mathcal{E}^{\mathsf{q}}_A &=& p_A^i \mathcal{E}_{ij} \Omega^j \, , \\ 
\mathcal{E}^{\mathsf{q}}_{AB} & = & 2 p_A^i p_B^j \mathcal{E}_{ab} + \tq_{AB} \mathcal{E}^{\mathsf{q}} \, , \\
\mathcal{B}^{\mathsf{q}}_A &=& p_A^i \varepsilon_{ijk} \Omega^j \mathcal{B}^k_{\; \; l} \Omega^l \, , \\
\mathcal{B}^{\mathsf{q}}_{AB} & = & 2 p_{(A}^i p_{B)}^l \varepsilon_{ijk} \Omega^j \mathcal{B}^k_{\; \; l} \, , \\
\mathcal{E}^{\mathsf{o}}  & = & \mathcal{E}_{ijk} \Omega^i \Omega^j \Omega^k \, , \\
\mathcal{E}^{\mathsf{o}} _A & = & p_A^i \mathcal{E}_{ijk} \Omega^j \Omega^k \, , \\
\mathcal{E}^{\mathsf{o}} _{AB} & = & 2 p_A^i p_B^j \mathcal{E}_{ijk} \Omega^k + \tq_{AB} \mathcal{E}^{\mathsf{o}} \, , \\
\mathcal{B}^{\mathsf{o}} _A & = & \frac{4}{3} p_A^i \varepsilon_{ijk} \Omega^i \mathcal{B}^k_{\; \; l m} \Omega^l \Omega^m \, , \, \mbox{ and} \\
\mathcal{B}^{\mathsf{o}} _{AB} & = & \frac{8}{3} p_{(A}^i p_{B)}^l \varepsilon_{ijk}  \Omega^j \mathcal{B}^k_{\; \; l m} \Omega^m  \, ,
\end{array} 
\ee
where 
\be
\Omega^i = [ \sin \theta \cos \phi, \sin \theta \sin \phi, \cos \theta] \label{Om}
\ee
is the standard radial vector in Euclidean $\Rbar^3$ and 
\bea
\mathcal{E}_{ijk} &=& \frac{1}{3} \left(D_i \mathcal{E}_{jk} + D_j \mathcal{E}_{ki} + D_k \mathcal{E}_{ij} \right) \, \mbox{ and} \\
\mathcal{B}_{ijk} & = & \frac{1}{8} \left(D_i \mathcal{B}_{jk} + D_j \mathcal{B}_{ki} + D_k \mathcal{B}_{ij} \,   \right) .
\eea

Each of these terms may be expanded in terms of the appropriate spherical harmonics, but for our purposes we note that direct integrations over the unit sphere with 
$d \Omega = \sin \theta d\theta d\phi$ give:
\bea
\int_{S^2} \mspace{-8mu} d \Omega \; {\mathcal{E}}^{qAB} {\mathcal{E}}^{\mathsf{q}}_{AB} &=& \left(\frac{32}{5} \pi \right) {\mathcal{E}}_{ij} {\mathcal{E}}^{ij}  \label{intFirst} \, , \\
\int_{S^2} \mspace{-8mu} d \Omega \;  \mathcal{B}^{qAB} \mathcal{B}^{\mathsf{q}}_{AB}  
&=&  \left( \frac{32}{5} \pi \right) \mathcal{B}_{ij} \mathcal{B}^{ij} \, ,  \\
\int_{S^2} \mspace{-8mu} d \Omega \; \mathcal{E}^{oAB} \mathcal{E}^{\mathsf{o}} _{AB}
& = & \left( \frac{32}{21} \pi \right) \mathcal{E}_{ijk} \mathcal{E}^{ijk}  \, , \mbox{and}  \\
\int_{S^2} \mspace{-8mu} d \Omega \; \mathcal{B}^{oAB} \mathcal{B}^{\mathsf{o}} _{AB} 
& = & \left( \frac{512}{189} \pi \right) \mathcal{B}_{ijk} \mathcal{B}^{ijk}  .
\eea

All other combinations are zero : 
\bea
\int_{S^2} \mspace{-8mu} d \Omega \; \mathcal{E}^{\mathsf{q} AB} \mathcal{B}^{\mathsf{q}}_{AB}  
=  \int_{S^2} \mspace{-8mu} d \Omega \; \mathcal{E}^{\mathsf{o} AB} \mathcal{B}^{\mathsf{o}} _{AB}  & = &   0  \, , \\
\int_{S^2} \mspace{-8mu} d \Omega \; \mathcal{E}^{\mathsf{q} AB} \mathcal{E}^{\mathsf{o}} _{AB}  =  
\int_{S^2} \mspace{-4mu} d \Omega \; \mathcal{B}^{\mathsf{q} AB} \mathcal{B}^{\mathsf{o}} _{AB}   & = & 0   \, , \mbox{and} \\
\int_{S^2} \mspace{-8mu} d \Omega \; \mathcal{E}^{\mathsf{q} AB} \mathcal{B}^\mathsf{o}_{AB}  =  
\int_{S^2} \mspace{-4mu} d \Omega \; \mathcal{E}^{\mathsf{o} AB} \mathcal{B}^{\mathsf{q}}_{AB} & = &  0 
\, . \label{intLast}
\eea

\section{Radial metric functions}
\label{appB}

The radial functions appearing in the tidal metric components (\ref{MetComp}) have the form : 
\be
\begin{array}{lll}
e_{1} & = & f^2 \\ 
e_{2} & = & f[1+\frac{1}{4x}(5+12 \ln x)-\frac{3}{4x^2}(9+4\ln x)+\frac{7}{4x^3} + \frac{3}{4x^4}] \\
e_{3} & = & f^2(1-\frac{1}{2x}) \\
e_{4} & = & f \\ 
e_{5} & = & f \left[1+\frac{1}{6x}\left(13+12 \ln x \right)-\frac{5}{2x^2} - \frac{3}{2x^3} - \frac{1}{2x^4} \right] \\ e_{6} & = & f \left(1 - \frac{2}{3x} \right) \\
e_{7} & = & 1 -\frac{1}{2x^2} \\ 
e_{8} & = & 1 + \frac{2}{5x}\left(4+3 \ln x \right) - \frac{9}{5x^2} -\frac{1}{x^3}\left(7+3 \ln x \right) + \frac{3}{5x^4} \\ 
e_{9} & = & f +\frac{1}{10 x^3} \\ 
b_{4} & = & f \\ 
b_{5} & = & f \left[1+\frac{1}{6x}\left(7+ 12 \ln x \right) - \frac{3}{2x^2} - \frac{1}{2x^3} - \frac{1}{6x^4}
\right] \\ 
b_{6} & = & f \left(1 - \frac{2}{3x} \right) \\ 
b_{7} & = & 1 -\frac{3}{2x^2} \\ 
b_{8} & = & 1+\frac{1}{5x}\left(5+6 \ln x\right) - \frac{9}{5x^2} - \frac{1}{5x^3}\left(2 + 3 \ln x\right) + \frac{1}{5x^4} \\ 
b_{9} & = & f - \frac{1}{10x^3} \, ,
\end{array}
\ee
where $x=r/(2M)$ and $f = 1-2M/r$. 
Their derivation is non-trivial and there is a (coordinate) 
gauge freedom in their final form. The reader is referred
to \cite{ericPRL} for details. Here we simply note that the freedom has been put to good use
to ensure that many of these functions vanish at $r=2 M$. In fact the only non-zero ones are 
$e_{7}=\frac{1}{2}$, $e_{9}=\frac{1}{10}$, $b_{7}=-\frac{1}{2}$ and $b_{9}=-\frac{1}{10}$. This greatly simplifies our calculations.

\end{appendix}

\end{document}